\documentclass[conference, a4paper]{IEEEtran}
\IEEEoverridecommandlockouts
\usepackage{cite}
\usepackage{amsmath,amssymb,amsfonts}
\usepackage{algorithmic}
\usepackage{graphicx}
\usepackage{textcomp}
\usepackage{xcolor}
\usepackage{bm}
\usepackage{pgfplots}
\usepackage{float}
\usepackage{flushend}
\usepackage{hyperref}
\usepackage[utf8]{inputenc}
\DeclareUnicodeCharacter{2212}{−}
\usepgfplotslibrary{groupplots,dateplot}
\usetikzlibrary{patterns,shapes,arrows,arrows.meta}
\pgfplotsset{compat=newest}
\usetikzlibrary{positioning}
\usetikzlibrary{calc}
\tikzset{
	-|-/.style={
		to path={
			(\tikztostart) -| ($(\tikztostart)!#1!(\tikztotarget)$) |- (\tikztotarget)
			\tikztonodes
		}
	},
	-|-/.default=0.5,
	|-|/.style={
		to path={
			(\tikztostart) |- ($(\tikztostart)!#1!(\tikztotarget)$) -| (\tikztotarget)
			\tikztonodes
		}
	},
	|-|/.default=0.5,
}

\makeatletter
\def\ps@IEEEtitlepagestyle{%
	\def\@oddfoot{\mycopyrightnotice}%
	\def\@oddhead{\hbox{}\@IEEEheaderstyle\leftmark\hfil\thepage}\relax
	\def\@evenhead{\@IEEEheaderstyle\thepage\hfil\leftmark\hbox{}}\relax
	\def\@evenfoot{}%
}
\def\mycopyrightnotice{%
	\begin{minipage}{\textwidth}
		\scriptsize
		\copyright 2021 IEEE.  Personal use of this material is permitted. Permission from
		IEEE must be obtained for all other uses, in any current or future
		media, including reprinting/republishing this material for advertising
		or promotional purposes, creating new collective works, for resale or
		redistribution to servers or lists, or reuse of any copyrighted
		component of this work in other works. DOI: \url{https://doi.org/10.1109/VCIP53242.2021.9675421}
	\end{minipage}
}
\makeatother

\graphicspath{{figures/}}

\def\BibTeX{{\rm B\kern-.05em{\sc i\kern-.025em b}\kern-.08em
    T\kern-.1667em\lower.7ex\hbox{E}\kern-.125emX}}

\DeclareMathOperator*{\argmax}{\arg\!\max}
\DeclareMathOperator*{\argmin}{\arg\!\min}

\newcommand{\fig}{Fig.\ }
\newcommand{\tab}{Table\ }

\newcommand{\LinearRegA}{a}
\newcommand{\LinearRegB}{b}
\newcommand{\NumRefImages}{N}
\newcommand{\NumBlocks}{M}
\newcommand{\BlockSize}{S}
\newcommand{\MaskedBarNMMean}{\bar{m}}
\newcommand{\ReferenceBarNMMean}{\bar{r}}
\newcommand{\BestReference}{z}

\newcommand{\MaskedBar}{\bm{m}}
\newcommand{\EFNeighbor}{\bm{n}}
\newcommand{\MaskedBarNM}{\bm{\tilde{m}}}
\newcommand{\CoordX}{\bm{x}}
\newcommand{\CoordY}{\bm{y}}

\newcommand{\Block}{\bm{B}}
\newcommand{\DistortedImage}{\bm{D}}
\newcommand{\CleanDistortedImage}{\bm{C}}
\newcommand{\IndicesBM}{\bm{I}}
\newcommand{\Mask}{\bm{M}}
\newcommand{\ReferenceBar}{\bm{R}}
\newcommand{\ReferenceBarNM}{\bm{\tilde{R}}}

\newcommand{\ReferenceImages}{\mathcal{R}}
\newcommand{\EFPixelSet}{\mathcal{P}}
\newcommand{\EFNeighborSet}{\mathcal{N}}

\newcommand{\T}{\text{T}}

\begin{document}

\title{Spatio-spectral Image Reconstruction Using Non-local Filtering}

\author{\IEEEauthorblockN{Frank Sippel, Jürgen Seiler, and André Kaup}
	\IEEEauthorblockA{\textit{Multimedia Communications and Signal Processing} \\
		\textit{Friedrich-Alexander University Erlangen-Nürnberg (FAU)}\\
		Cauerstr. 7, 91058 Erlangen, Germany \\
		\{frank.sippel, juergen.seiler, andre.kaup\} @fau.de}
}

\maketitle

\begin{abstract}
In many image processing tasks it occurs that pixels or blocks of pixels are missing or lost in only some channels. For example during defective transmissions of RGB images, it may happen that one or more blocks in one color channel are lost. Nearly all modern applications in image processing and transmission use at least three color channels, some of the applications employ even more bands, for example in the infrared and ultraviolet area of the light spectrum. Typically, only some pixels and blocks in a subset of color channels are distorted. Thus, other channels can be used to reconstruct the missing pixels, which is called spatio-spectral reconstruction. Current state-of-the-art methods purely rely on the local neighborhood, which works well for homogeneous regions. However, in high-frequency regions like edges or textures, these methods fail to properly model the relationship between color bands. Hence, this paper introduces non-local filtering for building a linear regression model that describes the inter-band relationship and is used to reconstruct the missing pixels. Our novel method is able to increase the PSNR on average by 2 dB and yields visually much more appealing images in high-frequency regions.
\end{abstract}

\section{Introduction}
\label{sec:intro}

\begin{figure}[t]
	\begin{center}
		\begin{tikzpicture}[scale=0.25]
			\node[inner sep=0pt] (rgb) at (0,4) {\includegraphics[width=.13\textwidth]{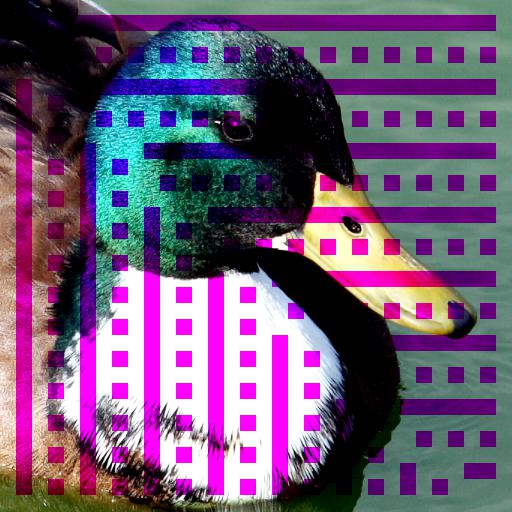}};
			
			\draw[-{Stealth[scale=2]}] (5, 4) -- (6.5, 4);
			\draw[-{Stealth[scale=2]}] (5, 5) -- (6.5, 6);
			\draw[-{Stealth[scale=2]}] (5, 3) -- (6.5, 2);
			
			\node[inner sep=0pt] (red) at (8.5,8) {\includegraphics[width=.05\textwidth]{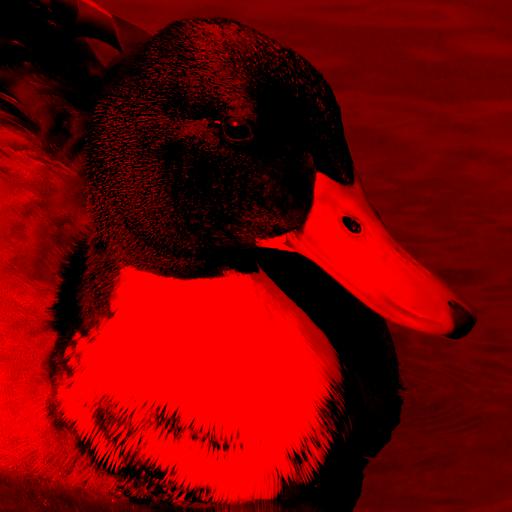}};
			\node[inner sep=0pt] (green) at (8.5,4) {\includegraphics[width=.05\textwidth]{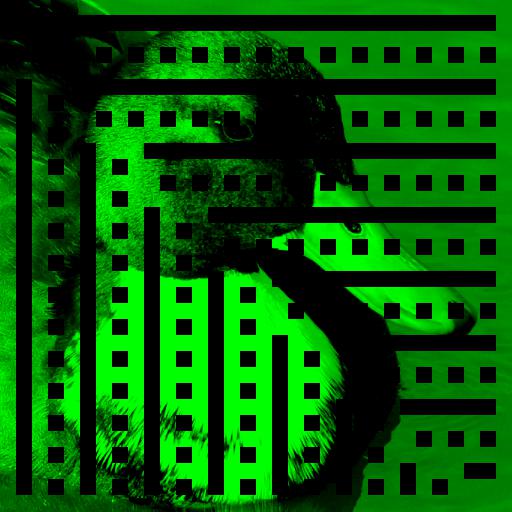}};
			\node[inner sep=0pt] (blue) at (8.5,0) {\includegraphics[width=.05\textwidth]{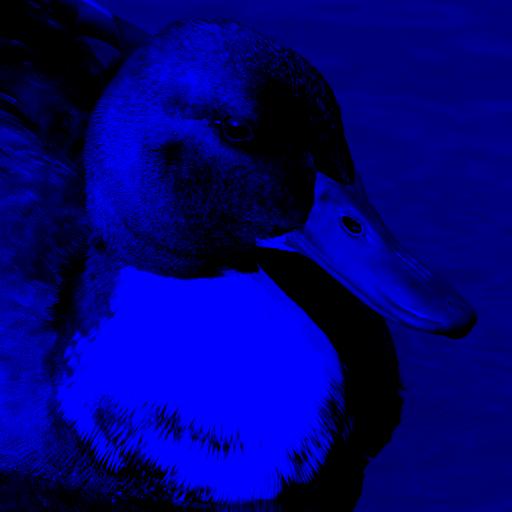}};
			
			\draw[-{Stealth[scale=2]}] (11.5, 4) -- (18, 4) node[midway, above] (reco) {Reconstruction};
			
			\node[inner sep=0pt] (rgb) at (24,4) {\includegraphics[width=.13\textwidth]{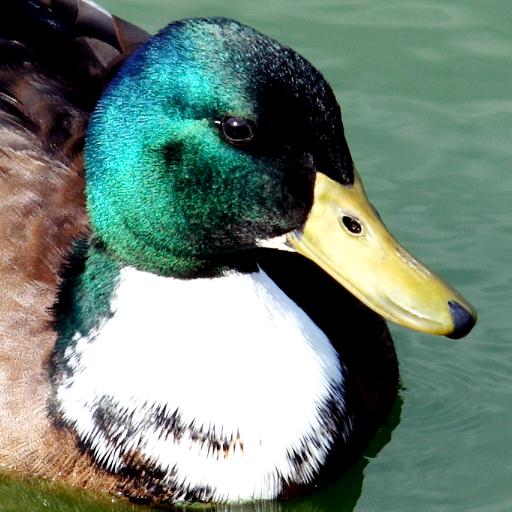}};

		\end{tikzpicture}
	\end{center}
	\caption{The addressed problem of this paper. The green channel has some missing pixels, while the red and the blue channel are complete. The red and the blue channel serve as reference channels to reconstruct the green channel.}
	\label{fig:problem_statement}
\end{figure}
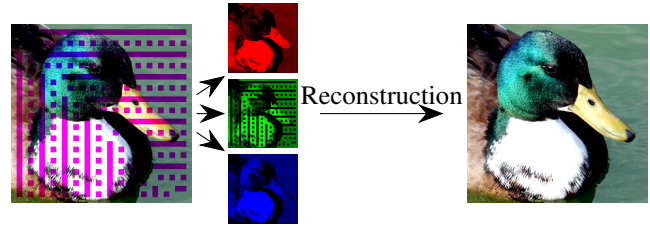

In many image processing tasks missing or defected pixels occur only in some of the available color channels. The goal of reconstruction algorithms is to extrapolate these missing pixels using spatial and spectral information. The setup for this paper is shown in \fig\ref{fig:problem_statement}. The green channel has a lot of missing pixels while the red and the blue channel are fully available. Thus, one can not only use the spatial information within the green channel, but the spectral and spatial information of the red and blue channel can also be exploited for reconstruction.

This situation often occurs when recovering missing blocks for defective transmissions of JPEG images \cite{skodras_jpeg_2001} and videos \cite{sullivan_overview_2012}, since the channels of the different blocks are quantized and transmitted individually. Another example is remote hyperspectral imaging \cite{bioucas-dias_hyperspectral_2013}, where only some bands contain defected pixels due to atmospheric effects. Finally, cross-spectral multi-view cameras like in \cite{genser_camera_2020}, where different views have to be warped to the center perspective, are also affected by missing pixels in some bands. Since every camera picks up slightly different viewing angles, occlusions occur. Hence, there are pixels that are recorded by the center camera, but are not visible to the peripheral cameras. Thus, in the end these pixels need to be reconstructed. Luckily, other cameras, especially the center camera, can see these pixels, however, these cameras pick up different parts of the light spectrum. Thus, using these reference channels, a model can be built to predict the value of the occluded pixels.

Generall, existing reconstruction algorithms can be classified into three categories. First, the problem of missing pixels and blocks can be tackled using only spatial information, e.g., \cite{zhai_bayesian_2010} uses discrete cosine pyramids, \cite{barnes_patchmatch_2009} and \cite{korman_coherency_2016} try to find similar patches within the image, \cite{sun_image_2005} first extrapolates structures and tries to find suitable patches afterwards, and \cite{tv} uses convex optimization and tries to minimize the total variation. Second, other algorithms also take the color information into account, for example \cite{fsr}, which builds a model based on spatial basis functions with different frequencies, or by utilizing a Bayesian approach \cite{parmar_bayesian_2007}. The third category are fully spatio-spectral reconstruction methods, which assume that the same spatial structures are visible in all color channels. Thus, these methods exploit spatio-spectral correlation. The state of the art for these type of algorithms is Content-Adaptive Selective Extrapolation (CASE) \cite{case}, which is based on estimating a local linear regression model and predicts the missing values within the block by applying this model. Unfortunately, due to the block-based processing approach, blocks with high frequency content are often modeled inaccurately due to the rapid change between different pixels for the linear color model. Hence, the idea of this paper is to employ a block-matching procedure to isolate the pixels that belong to the same homogeneous region and build the linear regression model based on these pixels.

\begin{figure*}[ht]
	\begin{center}
		\begin{tikzpicture}[scale=0.25]
			\input{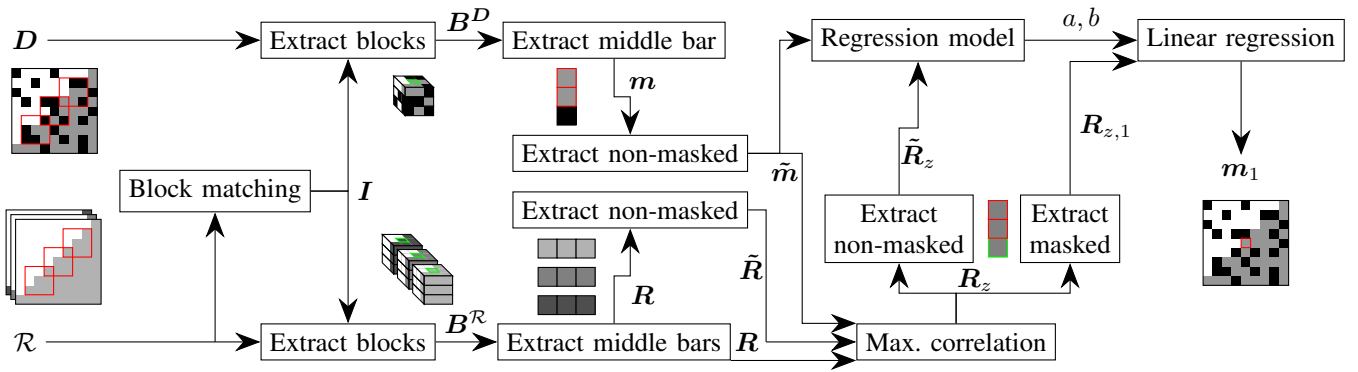}
		\end{tikzpicture}
	\end{center}
	\vspace*{-0.2cm}
	\caption{The basic concept of our novel method NOCS.}
	\label{fig:nocs_concept}
\end{figure*}

The goal of spatio-spectral image reconstruction is to estimate the missing pixels of a distorted grayscale channel $\DistortedImage$ using a set of $\NumRefImages$ reference channels $\ReferenceImages$, which contain the same scene but in different spectral bands. For example, a goal could be to reconstruct missing pixels of the green channel of an RGB image using the undistorted red and blue channel. The mask, that indicates which pixels to reconstruct, is denoted as $\Mask$. A zero in the mask $\Mask$ shows a pixel to reconstruct while a one marks pixels that contain information. Hence, the goal is to reconstruct the clean channel $\CleanDistortedImage$ out of $\DistortedImage = \CleanDistortedImage \odot \Mask$, where $\odot$ represents the element-wise multiplication.

\section{Proposed Method}
\label{sec:method}

Our novel method \underline{No}n-Local \underline{C}ross-\underline{S}pectral Reconstruction (NOCS) uses a block-matching procedure on the reference channels to find pixels that approximately show the same content. Block-matching and non-local filtering algorithms became very popular with the introduction of block-matching 3D filtering \cite{dabov_image_2007} and non-local means filtering \cite{buades_non-local_2011}, which denoise grayscale images. Afterwards, these pixel stacks are used to estimate the parameters of a linear regression model. This model is then used to predict the pixel value of the masked pixel for which the blocks are matched. This basic procedure is shown in \fig \ref{fig:nocs_concept}. In the following a more detailed description is given.

The first operation is a block-matching procedure using all reference channels. Therefore, it is necessary to extract blocks for every pixel. These blocks centered around the two-dimensional image coordinate $\CoordX$ are denoted as $\Block^{\ReferenceImages}(\CoordX)$ and of size $\BlockSize \times \BlockSize$.
The $l_2$-norm is used for calculating the distance
\begin{equation}
	\text{d}(\CoordX, \CoordY) = \sum_{i=1}^{\NumRefImages} ||\Block^{\ReferenceImages}_i(\CoordX) - \Block^{\ReferenceImages}_i(\CoordY)||_2, 
\end{equation}
between two image coordinates, where the subscript denotes the index of the reference channel. Thus, the distance between two pixels is the sum of distances between the extracted blocks over all reference channels.

Afterwards, the $\NumBlocks$ blocks with the smallest distances are sorted according to their distance $\text{d}(\CoordX, \CoordY)$ and the locations of these blocks are stored in the list $\IndicesBM_{\CoordX}$. Therefore, the coordinate $\CoordX$ itself with distance zero is always at the first position, i.e., $\IndicesBM_{\CoordX}(1) = \CoordX$. This list is then used to build reference bars $\ReferenceBar(\CoordX)$, where $\ReferenceBar(\CoordX)$ stores all reference bars as rows in a matrix, and the masked bar $\MaskedBar(\CoordX)$ as shown in \fig\ref{fig:nocs_concept}. Since the following procedure is the same and independent for every pixel and thus bars $\ReferenceBar(\CoordX)$ and $\MaskedBar(\CoordX)$ are abbreviated as $\ReferenceBar$ and $\MaskedBar$, respectively. Only bars, where a masked pixel is at the first index, are considered since this pixel is the only one that will be reconstructed in the bar.

Obviously, there are blocks that are easier to reconstruct since they have more valid pixels that can be used to build the model. Thus, the next steps are done in an iterative way. To speed up the reconstruction process, several bars are processed in parallel. A good measure how well the reconstruction can be done is the number of non-masked pixels in the masked bar. Therefore, all non-reconstructed bars are sorted for every iteration by the number of masked pixels in the bar in a descending order. Afterwards, the first 10\% of the sorted bars are processed in parallel.

For every pixel to reconstruct, similar pixels are found and the corresponding reference bars are set up. The next step is to build the linear regression model. This model is used to reconstruct the masked pixel in the bar. For this, first the best reference is found by maximizing the correlation between the non-masked pixels $\MaskedBarNM$ of the masked bar and the pixels $\ReferenceBarNM$ at the same locations in the reference bars. The reference index is found by
\begin{equation}
	 \BestReference = \argmax_i \frac{(\MaskedBarNM - \MaskedBarNMMean)^\T (\ReferenceBarNM_i - \ReferenceBarNMMean_i)}{||\MaskedBarNM - \MaskedBarNMMean||_2 \cdot ||\ReferenceBarNM_i - \ReferenceBarNMMean_i||_2},
\end{equation}
where $\ReferenceBarNM_i$ denotes the $i$-th row of the non-masked reference bar matrix $\ReferenceBarNM$, and $\MaskedBarNMMean$ and $\ReferenceBarNMMean_i$ denote the mean of $\MaskedBarNM$ and $\ReferenceBarNM_i$, respectively. Of course, if there is only one reference channel available, this procedure can be skipped and $\BestReference$ is set to $1$. Then, it is assumed that there is a linear relationship between the masked bar and the best reference bar. Thus, the model reads as 
\begin{equation}
	\MaskedBarNM = \LinearRegA \cdot \ReferenceBarNM_{\BestReference} + \LinearRegB,
\end{equation}
where $\LinearRegA$ and $\LinearRegB$ are the linear regression parameters. These parameters are then found by
\begin{equation}
	\hat{\LinearRegA}, \hat{\LinearRegB} = \argmin_{\LinearRegA, \LinearRegB} ||\LinearRegA \cdot \ReferenceBarNM_{\BestReference} + \LinearRegB - \MaskedBarNM||_2^2.
\end{equation}
Since this optimization problem is a simple minimization of a quadratic function, the parameters can be found in closed form
\begin{equation}
	\hat{\LinearRegA} = \frac{(\ReferenceBarNM_{\BestReference} - \ReferenceBarNMMean_{\BestReference})^\T (\MaskedBarNM - \MaskedBarNMMean)}{(\ReferenceBarNM_{\BestReference} - \ReferenceBarNMMean_{\BestReference})^\T(\ReferenceBarNM_{\BestReference} - \ReferenceBarNMMean_{\BestReference})} \quad \text{and} \quad \hat{\LinearRegB} = \MaskedBarNMMean - \hat{\LinearRegA} \cdot \ReferenceBarNMMean_{\BestReference}.
\end{equation}
These model parameters can then be used to reconstruct the masked pixel, i.e., the first pixel of the current bar
\begin{equation}
	\MaskedBar_1 = \hat{\LinearRegA} \cdot \ReferenceBar_{\BestReference, 1} + \hat{\LinearRegB},
\end{equation}
where the $1$ in the subscript denotes the first element of the vector. Thus, in the distorted channel, the according pixel can be set to the reconstructed value $\DistortedImage(\CoordX) = \MaskedBar_1$. Furthermore, the image mask at the same position is set to one $\Mask(\CoordX) = 1$.

It may occur that the block-matching matches purely masked blocks in a circular manner, i.e., a closed masked region where all of the matched pixels again fall into this region. Thus, the linear regression cannot be build since there are no non-masked pixels. A trivial, but not practical, example for this case would be if the distorted channel is completely masked. Hence, an emergency strategy is necessary, if there are only completely masked bars left in the iterative procedure. Out of the set of pixels in the closed region, only the set of pixels $\EFPixelSet$ with at least one non-masked neighbor is considered. Then, the squared difference in the reference channels between these pixels and the corresponding non-masked direct neighbors with pixel distance one is minimized. Thus, for each pixel $\CoordY$ from the set $\EFPixelSet$, every non-masked direction out of $\{\begin{pmatrix}1 & 0\end{pmatrix}^\T, \begin{pmatrix}0 & 1\end{pmatrix}^\T, \begin{pmatrix}-1 & 0\end{pmatrix}^\T, \begin{pmatrix}0 & -1\end{pmatrix}^\T\}$ is put into the corresponding neighbor direction set $\EFNeighborSet(\CoordY)$. Therefore, the minimization problem can be written as
\begin{equation}
	\hat{\CoordY}, \hat{\EFNeighbor} = \argmin_{\CoordY \in \EFPixelSet, \EFNeighbor \in \EFNeighborSet\left(\CoordY\right)}  \sum_{i=1}^{\NumRefImages} \left(\ReferenceImages_i(\CoordY) - \ReferenceImages_i(\CoordY + \EFNeighbor)\right)^2.
\end{equation}
This results in a pair of pixels with one masked pixel $\hat{\CoordY}$ and one non-masked pixel $\hat{\CoordY} + \hat{\EFNeighbor}$, which have the minimal difference between them in the reference channels. Then, the value of the non-masked pixel in the masked channel is copied to the position of the masked pixel $\DistortedImage(\hat{\CoordY}) = \DistortedImage(\hat{\CoordY} + \hat{\EFNeighbor})$. Afterwards, the emergency procedure is left and the normal algorithm can continue, since this pixel can be used for other masked pixels again. Of course, it may occur that this emergency mode is entered multiple times during the reconstruction procedure.

\section{Experimental results}
\label{sec:experiment}

\begin{figure}
	\begin{center}
		\input{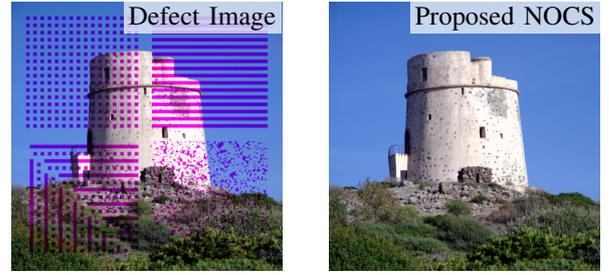}
	\end{center}
	\vspace*{-0.2cm}
	\caption{A masked image and the reconstructed image of the TECNICK dataset.}
	\label{fig:nice_image}
\end{figure}

The evaluation is done utilizing the TECNICK dataset \cite{asuni_testimages_2013}. This dataset contains 100 images of natural and urban scene with a resolution of $1200 \times 1200$ pixels. For the evaluation, The green channel of the RGB images was masked using a masking pattern as shown in \fig\ref{fig:nice_image}. The pattern itself is split into four different subpatterns. The top left pattern has uniformly distributed rectangles that mask the channel. Next to it on the right, horizontal bars are utilized as masking shape. In the bottom left, a mixture of rectangles, horizontal bars and vertical bars is used as mask. Finally, in the bottom right, rectangles are randomly distributed. These rectangles unmask the corresponding area. These different areas represent errors occuring in different applications. For example, block losses are occuring in a defective transmission, while the random pattern is a typical example for missing pixels for occlusion cases \cite{meisinger_spatiotemporal_2007}. Only the green channel is masked since one of the key features is that several reference channels can be taken into account. Thus, the only possibility for an RGB image to have several reference channels is to only mask one channel. Of course, it would also be possible to only have one reference channel and two distorted channels. On the right side, one can see the reconstruction result of NOCS.

The parameters of NOCS were optimized using the DIV2K dataset \cite{agustsson_ntire_2017}. Luckily, our novel method only has two direct parameters and one parameter to speed up processing time. First, the direct parameters of NOCS are the block size $\BlockSize$ and the number of stacked blocks in the cube $\NumBlocks$, which are set to $9$ and $44$, respectively. These parameters are found by a grid search on the DIV2K dataset with a similar mask in the range between $6$ and $16$ with step size $1$ for $\BlockSize$ and in the range between $28$ and $80$ with step size $4$ for $\NumBlocks$. Furthermore, to speed up processing time, the maximum search range for the block-matching procedure is set to $33$. Thus, similar blocks are searched in the rectangle spanned by $16$ pixels to the left, right, up, and down. The parameters of the competitors \cite{tv}, \cite{fsr}, and \cite{case} are set to the values given in the corresponding paper.

\begin{table}
	\renewcommand{\arraystretch}{1.5}
	\centering
	\vspace*{-0.2cm}
	\caption{PSNR and SSIM results over the TECNICK database of diverse reconstruction methods.}
	\begin{tabular}{c||c|c|c|c}
		& TV \cite{tv}	& FSR \cite{fsr}	& CASE \cite{case}	& Proposed NOCS		\\ \hline\hline
		PSNR	& 32.37 dB		& 34.49	dB			& 44.33	dB			& \textbf{46.33	dB}	\\
		SSIM	& 0.967			& 0.981				& 0.996				& \textbf{0.997}	\\
	\end{tabular}
	\label{tab:evaluation_result}
	\vspace*{-0.2cm}
\end{table}

\begin{figure}
	\begin{center}
		\input{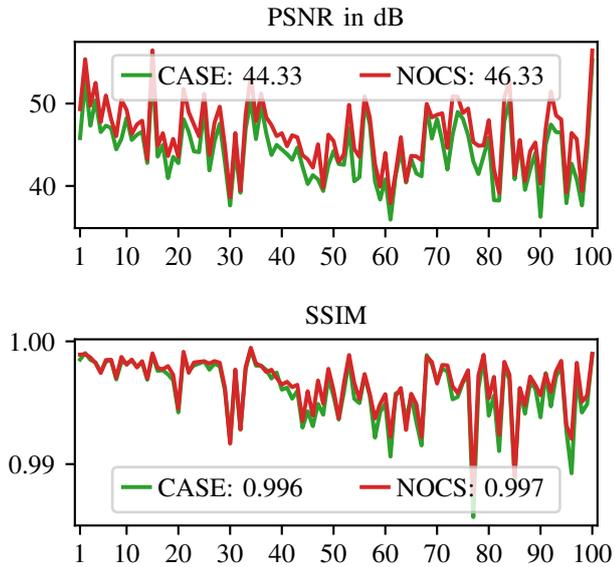}
	\end{center}
	
	\vspace*{-0.2cm}
	\caption{Evaluation of PSNR and SSIM over the whole TECHNIK database. The mean values of PSNR and SSIM of both methods are shown in the legend.}
	\label{fig:technick}
	\vspace*{-0.2cm}
\end{figure}

\tab\ref{tab:evaluation_result} shows the averaged PSNR and SSIM values for TV, FSR, CASE and our proposed NOCS. The methods that operate purely spatially TV and FSR are much worse than the ones that take spatio-spectral correlation into account, namely, CASE and NOCS. Quantitatively speaking, NOCS is more than 11 dB better than FSR and more nearly 14 dB better than the total variation-based approach. This proves that exploiting spatio-spectral correlation between channels is a promising technique to reconstruct missing pixels.

\fig\ref{fig:technick} shows the evaluation result for each of the 100 images of the TECNICK dataset for CASE and NOCS. Our proposed NOCS outperforms CASE in terms of SSIM with 0.997 to 0.996. Moreover, the average PSNR of NOCS is 2 dB higher than of CASE (44.3 dB vs. 46.3 dB). There are only 15 images where CASE has a slightly higher SSIM than NOCS. More importantly, there is not a single image that has a higher PSNR when reconstructed with CASE in comparison to NOCS.

\begin{figure}
	\begin{center}
		\input{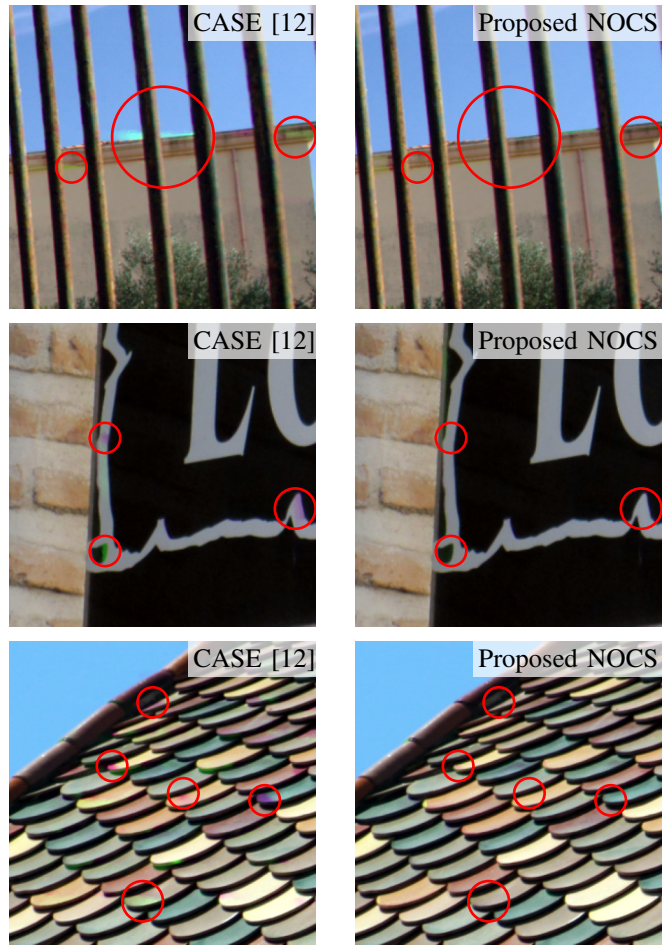}
	\end{center}
	
	\vspace*{-0.2cm}
	\caption{Various defects in high frequency areas highlighted using red circles of multiple images of the TECHNIK dataset. NOCS is able to remove nearly all of them. Best to be viewed enlarged on a monitor.}
	\label{fig:defects}
\end{figure}

\fig\ref{fig:defects} shows three image patches which verify that NOCS performs much better in high-frequency regions than CASE. In the first patch, a disadvantage of CASE is revealed. Due to the block-based processing approach, in high frequency areas the model is often build using a linear color model of a different texture, which leads to wrong reconstructions. Here, the linear regression model is setup using the building part of the image. As a consequence, the sky is wrongly reconstructed with this model. In the middle figure, it is shown that the non-local filtering does not only work for straight edges, but also for other shapes. Moreover, this example intuitively shows that CASE is often using the wrong model in regions with edges. For the red circle on the right, the model is built mainly using the black pixels. Thus, the linear regression parameters are close to zero. This leads to green values close to zeros and thus a purplish color. On the other hand, for the circle in the bottom right, the model is built mainly using white pixels, which leads to the greenish color. In the last example, the tiles frequently change their color. Thus, the regression model built by CASE is often invalid, which results in many different colors in masked areas that do not fit the color of the corresponding tile at all. NOCS handles this situation much better, especially since many tiles are repeated in the neighborhood.

\section{Conclusion}
\label{sec:conclusion}

A novel spatio-spectral reconstruction method called Non-Local Cross-Spectral Reconstruction (NOCS) was introduced. In comparison to the state-of-the-art algorithms, NOCS uses a block-matching procedure to find non-local similar pixels. These similar pixels are then utilized to build a linear regression model using only the non-masked pixels. After calculating the linear regression model parameters in closed-form, this model can be used to reconstruct the masked pixel for which the similar pixels are found. By exploiting the fact that same or similar pixels are distributed all over the image, e.g., along an edge, the model parameters can be much more precisely built in high-frequency areas than using traditional block-based processing. The evaluation showed that the novel method NOCS increases the PSNR on average by 2 dB, while not a single image out of the 100 test images was reconstructed worse by NOCS than when utilizing the current state of the art. As an outlook, we plan to remove the constraint that at least one channel has to be fully available. Furthermore, this will also allow to take channels with missing pixels into account for reconstructing pixels of other channels with missing pixels.

\vfill\pagebreak
\clearpage

\bibliographystyle{ieeetran}
\bibliography{refs}
\vspace{12pt}

\end{document}